# Performance assessment of a tightly baffled, long-legged divertor configuration in TCV with SOLPS-ITER

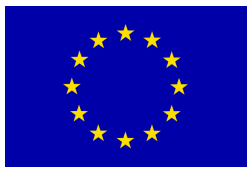

This work has been carried out within the framework of the EUROfusion Consortium and has received funding from the Euratom research and training programme 2014-2018 and 2019-2020 under grant agreement No 633053. The views and opinions expressed herein do not necessarily reflect those of the European Commission.

# Performance assessment of a tightly baffled, long-legged divertor configuration in TCV with SOLPS-ITER

G. Sun, H. Reimerdes, C. Theiler, B. P. Duval, M. Carpita, C. Colandrea, O. Février and the TCV team[1]

Ecole Polytechnique Fédérale de Lausanne (EPFL), Swiss Plasma Center (SPC), CH-1015 Lausanne, Switzerland

E-mail: guang-yu.sun@epfl.ch

**Abstract**

Numerical simulations explore a possible tightly baffled, long-legged divertor (TBLLD) concept in a future upgrade of the Tokamak à configuration variable (TCV). The SOLPS-ITER code package is used to compare the exhaust performance of several TBLLD configurations with results from unbaffled and baffled TCV configurations. The investigated TBLLDs feature a range of radial gaps between the separatrix and the divertor baffles, with a smaller gap resulting in tighter baffling. All modelled TBLLDs are predicted to lead to a denser and colder plasma in front of the targets and increase the power handling by factors of 2-3 compared to the present, baffled, divertor and by up to a factor of 12 compared to the original, unbaffled, configuration. This improved TBLLD performance is attributed to an increased neutral confinement with more plasma-neutral interactions in the divertor region. Both power handling capability and neutral confinement increase with tighter baffling. The core compatibility of TBLLDs with nitrogen seeding is also evaluated and the detachment window, with acceptable core pollution, for these TBLLDs is explored, showing a reduction of the required upstream impurity concentration to achieve detachment by up to 18% with tighter baffling.

## 1. Introduction

Heat exhaust remains a critical challenge for a fusion-energy reactor based upon the tokamak concept. As the scrape-off layer (SOL) width of the heat flux $\lambda_q$ scales inversely with the poloidal field rather than the reactor size [1], the unmitigated divertor heat flux of reactor devices such as DEMO is predicted to substantially exceed the target material limit [2]. This has motivated a range of studies on the optimization of divertor configurations, including radially extended divertor legs, higher-order magnetic nulls, additional secondary X-points, and increased divertor closure [3-5].

Increasing the connection length, in general, decreases the target temperatures, as highlighted by the two-point model [6]. An increased connection length can be achieved by extending the divertor leg vertically or radially, in the long-legged divertor [7] and super-X divertor [8], respectively, or by introducing an additional X-point above the divertor target (X-point target [9]) or just below the target (X divertor [10]). Numerical simulations have shown that poloidally extended divertor legs lead to remarkable increases of their power handling capabilities [7, 11-13].

---
[1] See the author list of H. Reimerdes et al. Nuclear Fusion 62 042018 (2022)

The poloidal extension of the divertor leg length, with the detachment front located away from the core plasma, is compatible with a high degree of divertor closure. Increased divertor closure yields higher neutral confinement that, in turn, enhances energy dissipation through plasma-neutral interactions, reducing the target heat load. In TCV, gas baffles of varied shape were installed to increase the divertor closure. They resulted in an increase in the divertor neutral pressure while decreasing divertor-core coupling and facilitating access to detachment [14-16], inline with design predictions [17]. Similar baffling approaches have also been applied to Alcator C-Mod [18]. Divertor closure can also be increased by target tilting [19-25], which reinforces the neutral recycling near the target and reduces neutral leakage towards the main plasma.

Inspired by a combination of poloidally extended divertor leg and divertor baffling, a tightly baffled, long-legged divertor (TBLLD) configuration is here proposed for possible future TCV upgrades. An initial implementation within TCV would seek to validate the concept experimentally. In the present work, numerical studies are reported to determine whether such an implementation could demonstrate key features of the TBLLD concept. This includes an increased power handling capability, increased neutral confinement, enhanced radial transport to the divertor baffles, and reduced nitrogen core concentrations. The study includes an initial optimization of the TBLLD geometry for TCV.

## 2. Divertor geometry and simulation setup

To take advantage of the existing TCV neutral beam heating systems [26, 27] the magnetic axis of considered configurations must be located close to TCV's mid-plane. Such configurations naturally feature a relatively long outer divertor leg that is ameanable to tight-baffling. Three outer leg widths between the outer strike point and the baffled outer leg are selected here for the TBLLD geometry for comparison with the unbaffled divertor and an existing TCV divertor configuration equipped with short-inner and long-outer (SILO) baffles,

Figure 1. The baffle structure is assumed to be cladded with the same carbon as the present vessel walls. The outer leg baffle contours are determined by the flux surfaces of the specific magnetic equilibrium, and the outer leg width is then scanned to explore the optimal divertor shape. For the existing TCV baffles, a trade-off between the plasma plugging and baffle recycling was found when varying the baffle size [15]. This is further tested for the combination of tight baffling and long-legged divertor. To facilitate this comparison, the strike point was kept at the same location, centered between the baffle side walls in the private flux region (PFR) and the common flux region (CFR). The distance between the strike point and either baffle is labeled $\Delta_B$. For characterization, the ratio of $\Delta_B$ and the local scrape-off layer width was chosen. The upstream outer mid-plane SOL heat flux width $\lambda_q$ is approximately 3mm in the present simulations, which is comparable with TCV measurements [28, 29]. The outer target poloidal flux expansion $f_{x,t}$ of ~5 in the present magnetic geometry must be taken into account in the evaluation of the ratio of $\Delta_B$ over $\lambda_q$. The parameters $\Delta_B$ and the ratio $\Delta_B/(f_{x,t} \cdot \lambda_q)$ of three TBLLDs, indexed as TBLL1, 2, 3 are shown in Table 1.

Table 1. Geometric parameters of the TBLLDs.

|  | TBLL1 | TBLL2 | TBLL3 |
|---|---|---|---|
| $\Delta_B$(mm) | 74.2 | 49.3 | 29.2 |
| $\Delta_B/(f_{x,t} \cdot \lambda_q)$ | 4.9 | 3.3 | 1.9 |

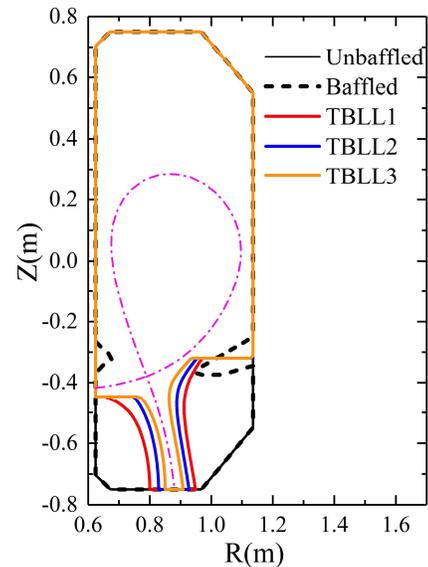

Figure 1. Considered TCV geometries with unbaffled and SILO-baffled divertor and three possible TBLLDs. The magenta curve represents the plasma separatrix.

The numerical simulations are performed with the SOLPS-ITER 3.0.7 code package, which combines the 2D fluid transport code B2.5 and the 3D kinetic neutral transport code EIRENE [30, 31]. Simulations are based on a typical lower single-null TCV equilibrium with a magnetic field of 1.4 T and a plasma current of 250 kA that was used in previous studies [32]. The employed magnetic equilibrium and an example of the B2.5 and EIRENE grids of TBLL2 are shown in Figure 2. Note that in the used SOLPS-ITER version the B2.5 mesh can only intersect the TCV contour at the targets. As a consequence, its radial extent is limited for the tightly baffled outer leg, Figure 2(b). Here, ions that cross the B2.5 mesh boundary are converted to neutrals. This description will be sensitive to the boundary conditions at the radial boundary of the B2.5 grid and can result in an unrealistic recycling distribution [33, 34], an effect that becomes more important as $\Delta_B$ is decreased. These limitations will be aleviated in the future by adopting the wide grid version of the SOLPS code package [35].

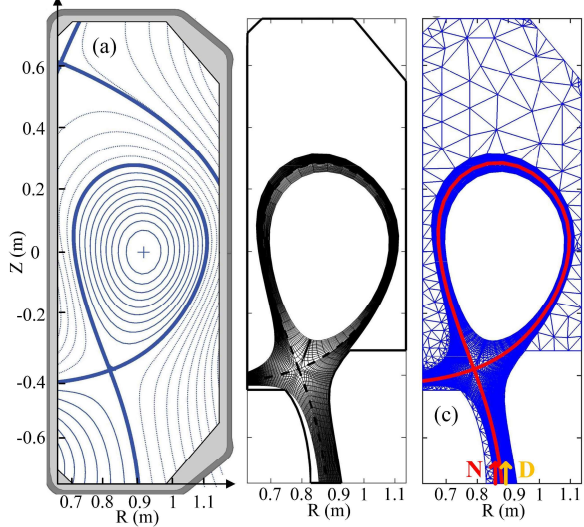

Figure 2. (a) Employed magnetic equilibrium (TCV discharge 64536). (b) B2.5 grid (c) EIRENE grid of TBLL2. Puffing locations are marked by arrows.

The considered species in the present simulation include deuterium, carbon and nitrogen. Deuterium is injected into the common flux region at the outer target and atomic nitrogen into the private flux region at the outer target, assuming a short mean free path for $N_2$ dissociation, Figure 2(c). Carbon is sourced by physical and chemical sputtering at the target, and from the chamber walls where only neutrals cause sputtering, at a chemical sputtering rate of 3.5%. Recycling coefficients are set to 0.99 for deuterium, 0 for carbon, 1 for neutral N and 0.3 for N ions. The considered reactions are listed in Table 2.

Table 2. Considered reactions in the simulations

| Reactions |
|---|
| $D + e \rightarrow D^+ + 2e$ |
| $D_2 + e \rightarrow D_2^+ + 2e$ |
| $D_2 + e \rightarrow D + D + e$ |
| $D_2 + e \rightarrow D^+ + D + 2e$ |
| $D_2 + D^+ \rightarrow D^+ + D_2$ |
| $D_2 + D^+ \rightarrow D_2^+ + D$ |
| $D_2^+ + e \rightarrow D^+ + D + e$ |
| $D_2^+ + e \rightarrow D^+ + D^+ + 2e$ |
| $D_2^+ + e \rightarrow D + D$ |
| $D^+ + e \rightarrow D$ |
| $D^+ + 2e \rightarrow D + e$ |
| $D^+ + D \rightarrow D + D^+$ |
| $C + e \rightarrow C^+ + 2e$ |
| $C + e \rightarrow C + h\nu$ |
| $D^+ + C \rightarrow D + C^+$ |
| $N + e \rightarrow N^+ + 2e$ |
| $N + e \rightarrow N + h\nu$ |

Cross-field transport coefficients $D_\perp = 0.2\text{m}^2\text{s}^{-1}$ and $\chi_{\perp,e} = \chi_{\perp,i} = 1.0\text{m}^2\text{s}^{-1}$ are taken and assumed constant. The choices of sputtering rate, recycling coefficients, and transport coefficients are consistent with previous TCV simulations using SOLPS-ITER and were in reasonable agreement with TCV L-mode experiments [36-38]. They may, therefore, overestimate the SOL width at high heating powers compatible with H-mode operation. The heating power is assumed to be equally shared between electrons and ions and will be varied in the following section to probe the power handling capability of TBLLDs. Far-SOL boundaries are constrained by constant radial density and temperature fall-off lengths that are adjusted to maintain an approximately flat fall-off length profile at the outer mid-plane. Drifts are deactivated in the present work. It should, however, be noted that the inclusion of drifts can influence the in-out divertor asymmetry, target parameter profiles, cross-field transport, etc., according to recent SOLPS simulations with drifts activated [37, 39-42]. The study of the effects of drifts in TBLLDs is left for the future.

## 3. Simulation results and analyses

The power handling capability of the divertor configurations is assessed by determining the maximum power that is compatible with detached divertor conditions in unseeded scenarios whilst monitoring the core nitrogen concentration in seeded scenarios. The detachment threshold is here defined, for simplicity, when the outer target peak electron temperature, $T_{ot,max}$, decreases to below 5eV.

Stable operation can also be bounded by the threshold for when the radiation front is displaced above the X-point and MARFEs occurs [43]. In SOLPS-ITER, this regime is usually numerically unstable and the present work, therefore, only discusses the low density side of the detachment window.

### 3.1. Power handling capability in unseeded plasmas

The power handling capability of the considered divertor geometries is evaluated by scanning the total power entering from the core boundary $P_{core}$, while keeping the plasma density at the outboard mid-plane separatrix constant at $1.5 \times 10^{19}\text{m}^{-3}$, Figure 3. While these scenarios are unseeded, erosion of carbon from the wall generates an intrisic impurity content. For any simulated value of $P_{core}$, all TBLLDs exhibit lower values of $T_{ot,max}$ than the baffled and unbaffled divertors. The maximum power that keeps the target detached, $P_{detached}^{max}$, determined from a linear interpolation of $T_{ot,max}$ in the simulations, increases from 103kW and 447kW for the unbaffled and baffled configurations to 832kW, 1105kW, and 1251kW for TBLL1, 2, 3, respectively. The TBLLDs, thereby, exceed the power handling capability of the unbaffled divertor by factors 8-12, and the baffled divertor by factors of 2-3. Note that, for $P_{core}$ above 1.3MW, TBLL3 yields a hotter outer target than TBLL2. This trend is unchanged by the addition of an inner baffle to reduce neutral leakage from the inner divertor. It should be underlined that predictions for TBLL3 may be more vulnerable to the limitations of the radial grid extent and the chosen transport coefficients. The upstream ionization rate tends to be overestimated with an over-narrow radial extent. Also the adopted wall boundary condition for the baffles should possibly be replaced by the sheath boundary condition for the narrow baffled outer leg. This phenomenon is not well understood and will be revisited in the future with the wide grid version of the SOLPS-ITER code, where baffles will not limit the plasma grid extension.

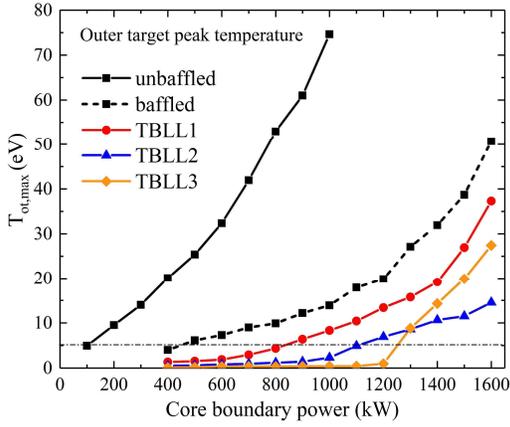

Figure 3. Dependence of the peak outer target electron temperature $T_{ot,max}$ on the simulated heating power $P_{core}$ for all considered divertor geometries with a constant upstream density of $1.5\times10^{19}$m$^{-3}$.

Target profiles for $P_{core}$=1MW are chosen to illustrate the divertor performance, Figure 4. At this power level, only TBLL2 and TBLL3 exhibit peak target temperature below 5eV, Figure 4(a). The TBLLDs significantly increase the target electron density, Figure 4(b), highest for TBLL3. The target particle flux profiles of all three TBLLDs are comparable with the baffled divertor, Figure 4(c), due to a larger reduction in the target temperature which counteracts the density increase. The TBLLDs show a lower target heat flux due to the lower target temperature and comparable target particle flux compared with the baffled divertor, Figure 4(d). Note that the power entering the outer divertor slightly varies between divertor geometries, even though the total core boundary power is the same, as will be discussed in section 3.3

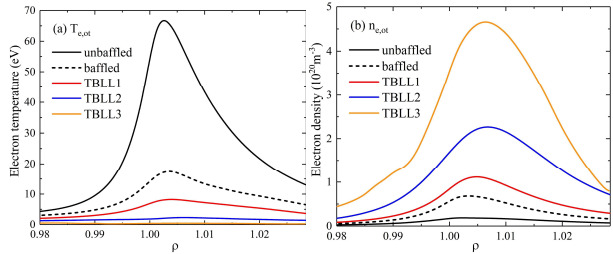
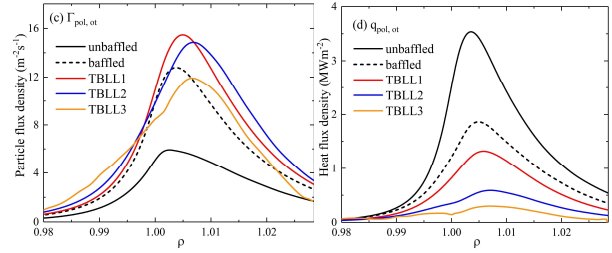

Figure 4. Outer target profiles of (a) electron temperature, (b) electron density, (c) poloidal plasma heat flux density and (d) poloidal particle flux density for all considered divertor geometries with 1MW core boundary power.

### 3.2. Detachment front and neutral confinement

Section 3.2 illustrates the increase of power handling capabilities introduced by the TBLLDs. When $P_{core}$ increases, the ionization region and detachment front are located closer to the hotter target. Here, the detachment front is defined as the poloidal position with peak temperature of 5eV. Note that the detachment front aligns with the ionization front, Figure 5, due to the significant drop of the ionization rate coefficient at temperatures below 5eV. The ionization front is defined as the poloidal position above which more than 90% of the total ionization in outer divertor region below the X-point occurs. At the same value of $P_{core}$, the front in the TBLLDs is located further from the target than in the baffled and unbaffled divertors, Figure 5. Decreasing $\Delta_B$ moves this front closer to the X-point, consistent with an increasing power handling capability to retain detachment. For the lowest simulated value of $P_{core}$ = 400kW, the front of TBLL3 is located at approximately half the distance between the X-point and outer target. All detachment locations are well below the X-point, i.e. far from any over-mitigation limit.

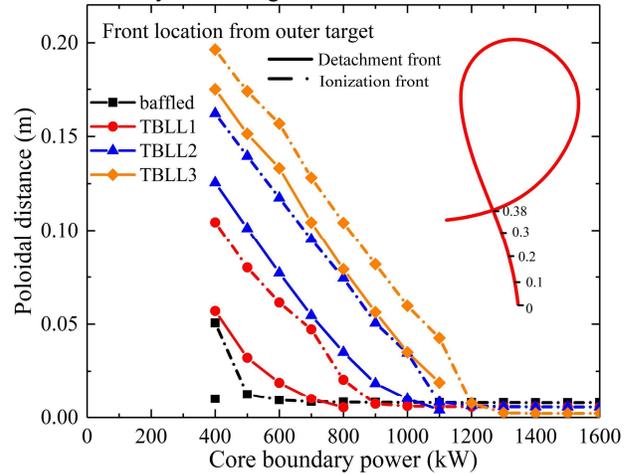

Figure 5. Core boundary power scan for the ionization and detachment front location from outer target. The poloidal distance from outer target at separatrix is marked in the subplot. Dash-dotted lines are ionization front locations and solid lines are detachment front locations. Unbaffled fronts are close to target and are not shown for better visibility.

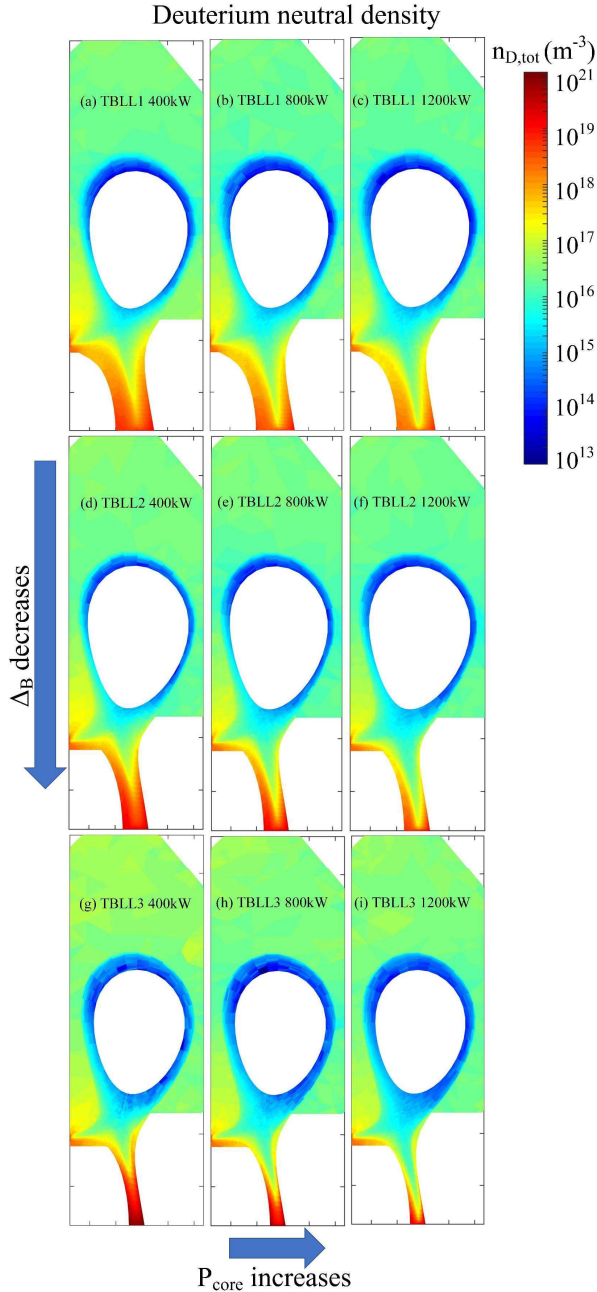

Figure 6. Total deuterium neutral density ($n_{D,tot} = n_D + 2n_{D2}$) distributions for TBLL1, 2 and 3 with increasing core boundary power from 400kW to 1200kW.

For an attached divertor, neutral leakage from the divertor throat is weak as the plasma density and temperature near the target are high, and the ionization mean free path, consequently, short. The divertor is, effectively, self-baffled by plasma plugging. When the divertor detaches, self-baffling weakens and recycling neutrals can traverse the divertor leg in the horizontal direction and, subsequently, bypass the plasma plug, reducing the neutral confinement. This channel can be mitigated by the TBLLD that tightly encompasses the plasma plug. Neutral density distributions for three TBLLDs for different $P_{core}$ are shown in Figure 6. For $P_{core}$ of 400kW and 800kW, all TBLLDs are detached (see Figure 3). At constant $P_{core}$, the outer divertor neutral density increases with decreasing $\Delta_B$ manifesting increased neutral confinement. As $P_{core}$ increases further, the dense neutral concentration divertor region shrinks and retreats towards the outer target and is eventually dissipated, signaling re-attachment. TBLL3 is the only divertor that remains detached for a $P_{core}$ of 1200 kW.

The minimum neutral density along the target is chosen to characterize the target neutral density $n_{D,ot}$, Figure 7. The target neutral density typically shows a "V" shaped profile, with the lowest density just outside the separatrix, where the electron temperature is highest. All three TBLLDs show a greater target neutral density $n_{D,ot}$, than the unbaffled and baffled configurations. At high power levels where all divertors are attached, the difference in target neutral densities becomes smaller. The significant increase in divertor neutral density for the TBLLD should lead to increased volumetric losses due to plasma-neutral interaction, discussed in section 3.3.

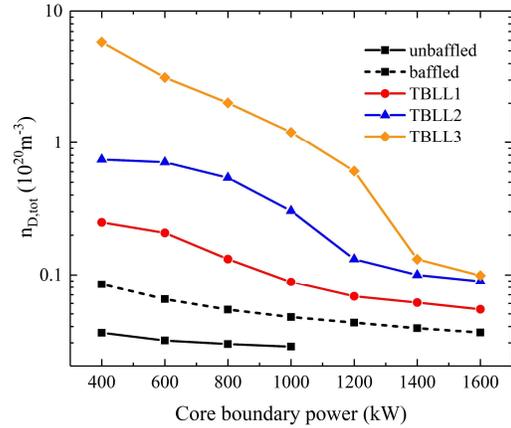

Figure 7. Dependence of the minimum total neutral density at the outer target $n_{D,tot}$ on core boundary power.

### 3.3. Power balance analysis

A power balance analysis of the outer divertor reveals the the effect of tight baffling on power dissipation mechanisms and helps explain the observed changes in the power handling capability and target parameters.

In a stationary divertor plasma, the power that enters the outer divertor is either lost via volumetric sinks or deposited by the plasma on the target or baffles,

$$Q_{ou} = Q_{vol} + Q_{ot} + Q_{baf} \qquad .(1)$$

$$Q_{vol} = \int_{OD} S_{tot} dV \qquad .(2)$$

where $Q_{ou}$ is the upstream power entering the outer divertor, $Q_{vol}$ the total volumetric losses to neutrals and radiation in the outer divertor region, $Q_{ot}$ the power the plasma deposits on the outer target and $Q_{baf}$ the power the plasma deposits on the the baffles. $Q_{vol}$ includes recombination, charge exchange, ionization, radiation and other energy losses due to plasma collisions with atomic and molecular neutrals.

The various contributions to the power balance have been extracted from the simulations with $P_{core} = 1MW$, Figure 8. In the unbaffled divertor (first bar) $Q_{ou}$=410kW enter the outer divertor of which $Q_{ot}$=247kW (corresponding to 60% of $Q_{ou}$) are received at the target, $Q_{vol}$=155kW lost in the volume and the $Q_{baf}$=1.5kW in the radial direction to the baffles. This corresponds to only a 1.5% discrepancy between the LHS and RHS of Equation (1).

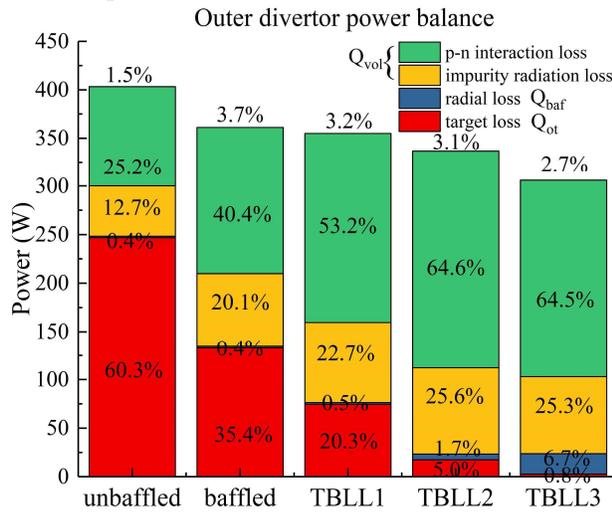

Figure 8. Power dissipation composition in the outer divertor for $P_{core} = 1$MW. The share of the total power loss is marked for all contributions, including plasma-neutral interaction, impurity radiation, radial power loss to the baffles and target power loss. The discrepancy between LHS and RHS of Equation (1) is given on the top of each bar.

In these unseeded scenarios, plasma-neutral interaction generally dominates over impurity radiation losses. The power loss due to plasma-neutral interaction and, to a lesser extent, from impurity radiation increases with decreasing $\Delta_B$, resulting in less power to the target, Figure 8. Note that for $P_{core} = 1$MW, only the TBLL2 and 3 are detached. In both cases, approximately 90% of the power entering the outer divertor is dissipated in the divertor volume, and the heat exhaust at the target is greatly reduced.

Radial plasma power losses to the baffles do not contribute significantly to the total power loss. Their contribution is stronger for the TBLLDs and increases as $\Delta_B$ decreases, but only reaching a maximum of 6.7% for TBLL3, Figure 8. Note that the employed SOLPS radial boundary conditions of constant density and temperature fall-off lengths may introduce systematic differences from a more realistic sheath boundary conditions at the TBLLD baffles.

Intrestingly, the power entering the outer divertor is observed to decrease with decreasing $\Delta_B$ but largely insignificant enough to explain the improved power exhaust characteristics of the TBLLDs. The present TBLLD geometries only baffle the outer divertor leaving the inner divertor unbaffled, possibly augmenting any inner divertor neutral leak. The treatment of inner divertor geometry will be further investigated in the future.

### 3.4 Compatibility with nitrogen seeding

Injecting radiative impurities to increase volumetric power loss is widely adopted to reduce the target heat exhaust and facilitate access to detachment and is currently thought to be mandatory in any reactor scenario. The choice of impurity species is, among other things, based on the reactor dimension and plasma conditions as the impurity-radiated power depends upon the SOL electron temperature. Nitrogen is widely used as the species for impurity seeding in TCV, and both L-mode and H-mode detachment have been achieved with nitrogen seeding [16, 44, 45].

Impurity ions generated beyond the stagnation point in the poloidal impurity velocity, determined by parallel force balance and E×B drifts, are transported upstream and may reduce the core performance by excessive radiative core emission [46, 47]. When the seeding rate increases, the divertor cools and the impurity ionization front displaces from the target towards the X-point, which also enhances the impurity leakage upstream, increasing core pollution [48].

Impurity transport is also affected the various shapes and sizes of baffles in a complex manner. Baffles cool the divertor, displacing the impurity ionization front towards the X-point and increase leakage. Baffles, simultaneoustly, improve impurity neutral confinement similarly to deuterium neutrals. The following analysis will evaluate the TBLLD detachment window with acceptable core impurity pollution levels.

The seeding rate is increased whilst maintaining an upstream density of $1.5 \times 10^{19}$m$^{-3}$ and a core boundary power of 1MW or 1.4MW. The impurity concentration at the outer mid-plane separatrix is calculated as the ratio of the carbon and nitrogen ion density to the electron density. Achieving cooler target temperatures generally requires more seeding and leads to a higher upstream impurity concentration, Figure 9. The detachment threshold, with peak outer target temperature below 5eV, is marked out. There is, however, no clear definition of "a" maximum impurity concentration at the outer mid-plane in TCV, but a lower concentration generally thought to improve the compatibility of the exhaust solution with core performance.

The operation window for seeded detachment requires that a target solution exists for peak target temperatures below 5eV with acceptable upstream impurity pollution. For $P_{core}$ of 1MW, TBLL2 and TBLL3 are detached for all seeding levels including the unseeded simulations, whereas TBLL1 is detached only for seeding rates above $4\times10^{20}$s$^{-1}$, and no detachment solution exists for baffled and unbaffled divertors for this range of seeding rates. The impurity concentrations in the outer mid-plane for all three TBLLDs are similar at a level ≤ 10% corresponding to a $Z_{eff}$ of ~ 3.3 at the maximum seeding rate. The core impurity concentration due to carbon without nitrogen seeding is between 7.6% and 8.6% for the TBLLDs. Note that the outer mid-plane impurity concentration is evaluated at a single point and its value can be slightly higher than the averaged core region concentration measured experimentally. Carbon density decreases with increasing nitrogen seeding rate due to a reducing target particle flux (note: chemical sputtering by ions at the target is found to be the main carbon source in SOLPS simulations of TCV). The total impurity concentration and core radiation factor increase with seeding rate.

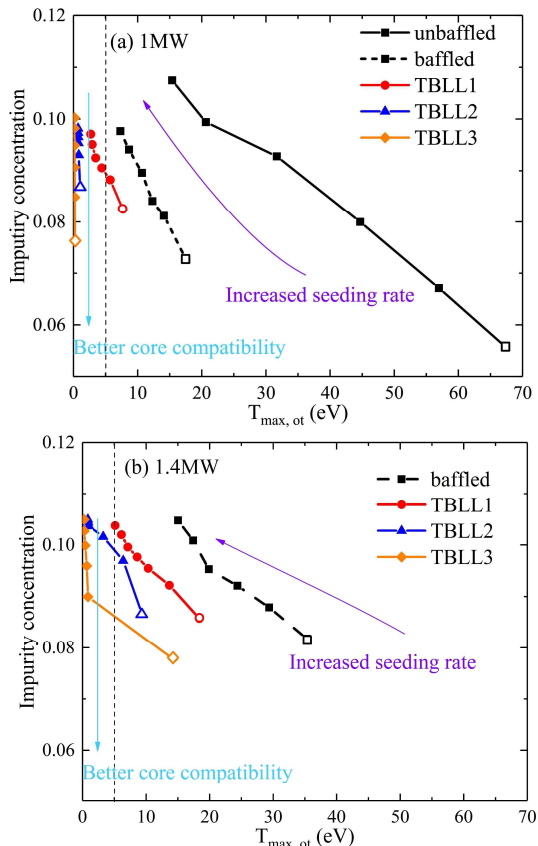

Figure 9. Dependence of the outboard mid-plane separatrix impurity concentration on the maximum outer target temperature for nitrogen seeding rates from 0 to $10^{21}$ s$^{-1}$, for the different divertor geometries with a core boundary power of (a) 1MW and (b) 1.4MW. The proxy for the detachment threshold is marked by the dashed line. Zero seeding points are marked with open symbols.

The impurity concentrations of all divertor geometries with 1.4MW core boundary power are higher than with 1MW for the same outer target temperature, as more impurities are needed to detach the divertor with higher input power. The impurity concentration at detachment threshold decreases with decreasing $\Delta_B$, up to 18% from the TBLL1 to the TBLL3. The presented seeding scan suggests that TBLLDs can reduce the impurity leakage to upstream so that higher divertor impurity concentration can be accommodated. This provides a broader operation window for nitrogen-seeded detachment. These effects become stronger with tighter baffling.

## 5. Conclusions

The performance of tightly baffled, long-legged divertors in TCV is evaluated with SOLPS-ITER simulation by comparisons with existing baffled and unbaffled TCV divertors. The radial gap between the separatrix and the baffle side walls, $\Delta_B$, is varied to optimize the TBLLD geometry of the outer divertor. The power handling capability and neutral confinement in the outer leg region are both significantly improved by all three modelled TBLLDs and increase with decreasing $\Delta_B$. The power balance analysis indicates that the increase in the exhaust losses is primarily due to increased volumetric losses in the divertor including both plasma-neutral interaction and impurity radiation, rather than an increase in radial plasma power losses to the baffles. The compatibility with nitrogen seeding is also tested, showing workable detachment windows within acceptable core pollution levels for the proposed TBLLDs. The present modeling results provide further motivation for the planned experimental test of the TBLLD concept on TCV, with the goal of developping improved power exhaust capabilities for fusion reactors.

## Acknowledgements

This work has been carried out within the framework of the EUROfusion Consortium, via the Euratom Research and Training Programme (Grant Agreement No 101052200 — EUROfusion) and funded by the Swiss State Secretariat for Education, Research and Innovation (SERI). Views and opinions expressed are however those of the author(s) only and do not necessarily reflect those of the European Union, the European Commission, or SERI. Neither the European Union nor the European Commission nor SERI can be held responsible for them. This work was supported in part by the Swiss National Science Foundation.